\documentclass[aps,prl,reprint, superscriptaddress,amsmath,amssymb,floatfix]{revtex4-2}

\usepackage{graphicx}
\usepackage{bm}
\usepackage{hyperref}
  \hypersetup{
      unicode=false, 
      pdffitwindow=false, 
      pdfstartview={FitH}, 
      pdfauthor={}, 
      colorlinks=true, 
      linkcolor=blue,        
      citecolor=blue,        %
      urlcolor = blue        %
  }
\usepackage[all]{hypcap}


\newcommand{\dsub}[1]{_{\text{#1}}}
\newcommand{\dchg}[1]{^{#1}}
\newcommand{\VZr}{\text{V}\dsub{Zr}}
\newcommand{\VO}{\text{V}\dsub{O}}
\newcommand{\VM}{\text{V}\dsub{M}}
\newcommand{\VTi}{\text{V}\dsub{Ti}}
\newcommand{\YZr}{\text{Y}\dsub{Zr}}
\newcommand{\ZrO}{\text{Zr}\dsub{O}}
\newcommand{\Zri}{\text{Zr}_i}
\newcommand{\trimer}{(\YZr\VO\YZr)}
\newcommand{\pair}{(\YZr\VO)}
\newcommand{\Ef}{\ensuremath{E_\text{F}}}
\newcommand{\Eform}{E^\text{f}}
\newcommand{\kB}{k_{\text{B}}}

\begin{document}

\title{Fermi-level mediated acceleration of flash sintering of oxide ceramics}

\author{Qin-Kun Li}
\affiliation{Department of Materials Science and NanoEngineering, Rice University, Houston, TX~77005, United States}

\author{Evgeni S. Penev}
\affiliation{Department of Materials Science and NanoEngineering, Rice University, Houston, TX~77005, United States}

\author{Boris I. Yakobson}\email{biy@rice.edu}
\affiliation{Department of Materials Science and NanoEngineering, Rice University, Houston, TX~77005, United States}
\affiliation{Department of Chemistry, Rice University, Houston, Texas 77005, United States}


\begin{abstract}
The atomistic understanding of flash sintering (FS) remains speculative, despite its efficiency and versatility in materials processing. Employing first-principles calculations we demonstrate how charge compensation of a range of defects in the prototypical Y-stabilized cubic ZrO$_2$ (YSZ) shifts Fermi level \Ef\ up during FS, thereby accelerating cation migration for fast mass transport. The charge transition of Zr vacancy, $\VZr\dchg{q}$, reduces its bulk diffusion barrier in $\VZr\dchg{-4}$ during flash by 2~eV, relative to $\VZr\dchg{0}$ before flash, which is triggered by the charge equilibrium of nonstoichiometric defects. The substituent defect $\YZr$, released by annihilating O vacancy, $\VO$, in $\YZr\VO\YZr$ defect complex, acts as electron acceptor and favors $\VZr\dchg{0}$ before flash whereas excess $\VO$, as electron donor thermally generated at the FS onset, upshift $\Ef$ and thus support $\VZr\dchg{-4}$. The proposed mechanism of Fermi-level mediated cation diffusion for YSZ is generalized to other flash-sintered ceramics and has considerable bearing on the general theory of FS techniques in oxide ceramics.
\end{abstract}

\maketitle

``Flash'' processing is a powerful versatile method for quick synthesis of a wide variety of materials, which is inaccessible by any other conventional methods~\cite{wang20,luong20}. This ultrafast process can swiftly realize compositional sintering in \emph{seconds} with ultrahigh heating rate and is an ideal option for high-throughput materials screening for ceramic-based new solid-state electrolytes that are critical for efficient and safe energy
storage~\cite{wang20,zheng23}. Such merits overshadow the centuries-old \emph{conventional}
sintering that has been persistently impeded by the hours processing time~\cite{wang20,luong20,rabinovich85}. The recently developed flash sintering (FS)
of functional metal oxide ceramics is one of the best examples, which has been first applied to sinter YSZ to full density in just a few seconds~\cite{cologna10}, and extensively applied to other metal oxides, such as ZnO, TiO$_2$, Al$_2$O$_3$, SrTiO$_3$, etc~\cite{biesuz19}. The similarity in flash-processing conditions underlies a possibly common mechanism, yet the reaction pathways and the underlying physics remain speculative.

YSZ is a popular ionic ceramic electrolyte for solid-oxide fuel cells~\cite{ch4_2016}; as a solid solution, (Y$_2$O$_3$)$_y$(ZrO$_2$)$_{1-y}$, the stoichiometric charge-compensating oxygen vacancies of concentration $[\VO]_{y/2}$, introduced by aliovalent substituent Y on Zr, $\YZr$, enhance its ionic (O$^{2-}$) conductivity and stability under redox conditions~\cite{ch4_2016}. We take YSZ as an example to elucidate some unique features of FS experiments during its three stages~\cite{cologna10,biesuz19,yadav17}. Stage I (incubation period) starts with an incipient rise of specimen conductivity under constant voltage. The flash event during stage II occurs with the abrupt rise of flash current and conductivity, signaled by bright light emission, the so-called ``flash''. At last, stage III is held in a quasi-steady state under constant current.

FS is distinguished by the current flowing in the specimen compared to other sintering
methods, which presumably stems from a range of
defects/impurities~\cite{cologna10,yadav17,raj11,jo24}. For a real material in a
well-defined thermodynamic state, the Fermi level \Ef\ explicitly determines the key
defect properties~\cite{freysoldt14,zunger21,park18,klein23,bonkowski24,wang25},
e.g., formation energy, equilibrium concentrations of various charge states, and especially
the dominant charge state regulated by the position of charge transition level relative to
\Ef. Conversely, defects act as charge acceptors or donors via optical/thermal ionization,
and the charge compensation ensures the charge neutrality of all charged species (ionized
defects and free charge carriers) in the lattice, thereby implicitly defining \Ef\ and
electronic properties of the material itself~\cite{freysoldt14,zunger21,park18,klein23}.
In general, donor (acceptor) defect doping donates electrons (holes) to the lattice and
results in an upward (downward) shift of \Ef. However, the role of defects in
understanding FS is still
elusive~\cite{cologna10,yadav17,jo24,majidi15,vendrell19,jo20,francis13}.
During stage I, the observed $p$-type electronic conductivity is attributed to Y
substitution and localized holes on O atoms~\cite{vendrell19,jo20}, the acceptor nature
of this defect possibly accounts for a \Ef\ closer to the valence band. At the FS onset
when specimen densification and ``flash'' occurs, the surging conductivity suggests a
significant rise of defect concentrations~\cite{cologna10,jo24,klein23,francis13},
which ideally originates from \Ef\ shift. As for the controversial sources of ``flash'',
thermal radiation~\cite{todd15} versus electroluminescence by electronic
transitions~\cite{yadav17}, defects matter in the latter mechanism but not in former.
Collectively, such observations raise the question as to how these defect-related phenomena
are connected with the microscopic mechanism of FS, especially of the heavier Zr ions
migration; mechanistic questions and quandaries about mass transport in FS still remain.

Here we employ first-principles calculations to unravel how charge compensation and
equilibrium of a range of defects in the prototypical YSZ lead to Fermi level shifts at
different stages of flash, thereby accelerating cation migration for fast mass transport
and sintering kinetics. The charge transition of $\VZr\dchg{q}$ reduces its bulk diffusion
barrier in $\VZr\dchg{-4}$ state during flash by 2~eV, compared to $\VZr\dchg{0}$ before
the surging of flash current. The $\VZr\dchg{q}$ charge transition is triggered by the
annihilation and dissociation of $\YZr\VO\YZr$ defect complex: O$_2$ inhalation annihilates
$\VO$ in the defect complex and releases $\YZr$ that serves as electron acceptor, and
$\VZr\dchg{0}$ is favored before flash, whereas at the FS onset, excess $\VO$ that are
thermally generated by O$_2$ exhalation upshifts Fermi level and thus support
$\VZr\dchg{-4}$, by contributing thermally excited electrons to form self-trapped electron
polaron on lattice Zr and to the conduction band edge. The proposed mechanism of
Fermi-level mediated cation diffusion scheme for YSZ is generalized to other flash sintered
ceramics and has considerable bearing on the general theory of flash sintering techniques
in oxide ceramics and other functional materials.

\begin{figure}[!tb]
  \includegraphics[width=\columnwidth]{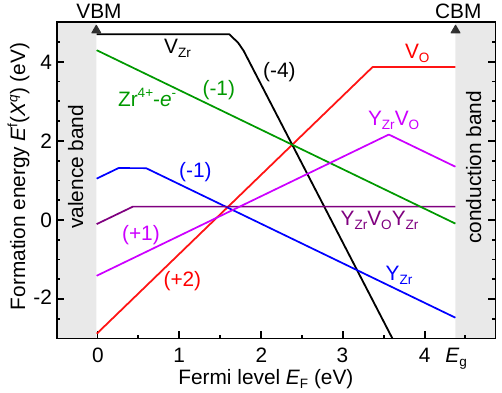}
  \caption{\label{fig:1}%
    Formation energy $\Eform(X\dchg{q})$ versus $\Ef$ for point defects $X\dchg{q}$. Only
    line segments with the lowest formation energy for different charge states are shown.}
\end{figure}

\textit{Energetics of relevant point defects}---Cubic ZrO$_2$ ($Fm\bar{3}m$) is the most stable of the three zirconia polymorphs (monoclinic, tetragonal and cubic) at temperature higher than $\simeq 2650$~K~\cite{shin18}, and can be stabilized down to room temperature by $\gtrsim 8$~mol\% Y$_2$O$_3$ doping (8YSZ). In the present thermodynamic treatment this material is represented by the nominal cation composition Y$_x$Zr$_{1-x}$O$_{2-x/2}$ with $x = 0.16$.

To unveil the underlying mechanism of FS, we calculate the formation energies of the
relevant point defects in cubic ZrO$_2$, shown in Fig.~\ref{fig:1}. Under given
thermodynamic conditions, the chemical potential $\Delta\mu$ of each element in YSZ can be
determined by the host stability condition,
\begin{equation}
  x \Delta\mu_\text{Y} + (1-x) \Delta\mu_\text{Zr}
  + (2-x/2)\,\Delta\mu_\text{O} = \Delta H_f(8\text{YSZ}),
  \label{eq:1}
\end{equation}
the fixed Y concentration that includes all Y-containing species (as discussed later),
\begin{equation}
  \sum_q [\YZr\dchg{q}] + 2\sum_q [\trimer\dchg{q}] + \sum_q [\pair\dchg{q}]
  = 0.16/\Omega,
  \label{eq:2}
\end{equation}
where $\Omega$ is the volume per formula unit of cubic ZrO$_2$, and the charge neutrality
condition~\cite{ma11,buckeridge19,zhang91},
\begin{equation}
  \sum_i q_i [X_i^{q_i+}] + [h^+] = \sum_j q_j [X_j^{q_j-}] + [e^-].
  \label{eq:3}
\end{equation}
For a given oxygen reservoir, i.e., fixed $\Delta\mu_\text{O}(T,p)$, and fixed 8YSZ
composition, eqs.~(\ref{eq:1})--(\ref{eq:3}), together with the defect thermodynamics
summarized in Supporting Information (SI), Sec.~S2, determine $\Delta\mu_\text{Y}$, $\Delta\mu_\text{Zr}$, \Ef, and the corresponding equilibrium defect and carrier concentrations self-consistently.

The defect formation energies $\Eform(X\dchg{q})$ allow one to determine the most favorable defect charge states $X\dchg{q}$, as well as the $q \to q'$ charge transition levels $\varepsilon(q/q')$ with respect to bandgap edges. Hereafter, energy positions are referenced to valence band maximum (VBM), hence the conduction band minimum (CBM) is at the bandgap $E_g$. Spin-polarized DFT calculations are performed for various defects $X\dchg{q}$ in charge states $q$, using the PBE functional~\cite{perdew96} along with on-site Hubbard $U$ correction~\cite{cococcioni05,dudarev98} to better capture the defect electronic behavior~\cite{lany09,falletta22} and reduce self-interaction error in standard DFT (see SI, Secs.~S1 and S2 for more details).

\begin{figure*}[t]
  \includegraphics[width=\textwidth]{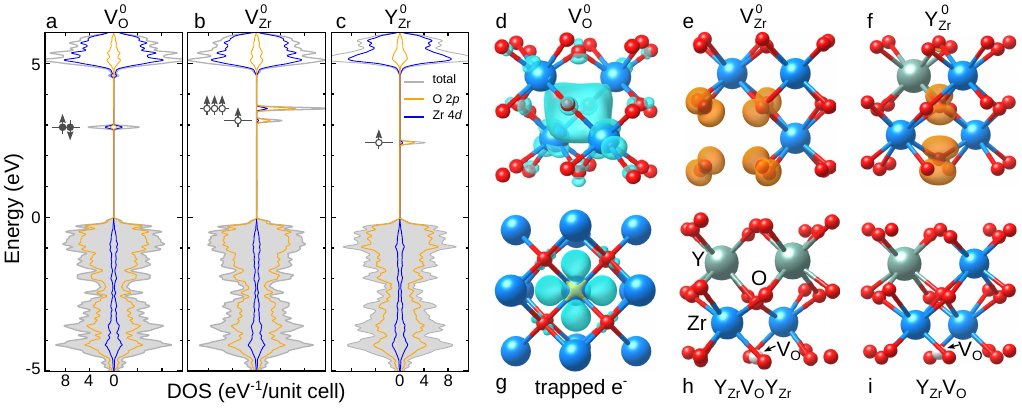}
  \caption{\label{fig:2}%
    Projected density of states (DOS) (right: spin-$\uparrow$, left: spin-$\downarrow$) of
    individual neutral (a) $\VO\dchg{0}$, (b) $\VZr\dchg{0}$, and (c) $\YZr\dchg{0}$. The
    localized gap states are magnified 10-fold; their electronic/spin configuration
    ($\bullet$ -- electron, $\circ$ -- hole) is schematically shown. The corresponding
    spin-resolved charge density isosurfaces (orange/cyan; 5\% of the respective maximum
    density) and local lattice structures are shown in (d), (e) and (f), respectively.
    (g) A self-trapped electron polaron localized on the Zr lattice, Zr$^{4+}$-$e^-$. The atomic structure of the $\trimer\dchg{0}$ and $\pair\dchg{+1}$ complexes are shown in (h) and (i), respectively. The white sphere is $\VO$, the second nearest neighbor to the $\YZr$.
    }
\end{figure*}

Yttrium substitution on Zr, $\YZr$ ($\Eform(\YZr\dchg{q})$ for $q = 0, \pm 1$ are shown in Fig.~\ref{fig:1}), and charge-compensating $\VO$ are the key, abundant defects in YSZ~\cite{peko13}. Figure~\ref{fig:2} illustrates the electronic structure of representative neutral defect states. The neutral $\YZr\dchg{0}$ features a localized hole state (corresponding in-gap peak in Fig.~\ref{fig:2}c) mostly on the second-nearest and
spatially extended to nearest neighbor O atoms (Fig.~\ref{fig:2}f), which asymmetrically
distorts the nearby Y/Zr-O bonds by 9--15\%. A negatively charged $\YZr\dchg{-1}$ almost
restores these bonds compared to pristine ZrO$_2$. The charge transition level $\varepsilon(0/-1) = 0.59$~eV indicates its shallow acceptor character when ZrO$_2$ FS occurs ($T \sim 1000$~K). Surprisingly, $\YZr$ can stabilize a second hole on another O atom and stay in $+1$ state, $\YZr\dchg{+1}$, if $\Ef < \varepsilon(+1/0) = 0.26$~eV; thus, $\YZr\dchg{+1}$ is a spin-one deep donor. In the entire range of $\Ef$, $\YZr$ acts as an
amphoteric substitution.

The neutral $\VO\dchg{0}$ is the most stable configuration of $\VO$ for $\Ef$ within 1~eV
of CBM; two electrons are trapped at the vacancy (Fig.~\ref{fig:2}d) and occupy in-gap
defect states (Fig.~\ref{fig:2}a), while single ionized $\VO\dchg{+1}$ is always thermodynamically unfavorable compared to $\VO\dchg{0}$ and $\VO\dchg{+2}$. As $\VO\dchg{+2}$ is the dominant charge state for $\Ef < \varepsilon(+2/0) = 3.36$~eV, the defect donates two electrons to the conduction band and thus acts as a persistent double donor and compensating center for acceptor-alike defects such as $\YZr$ and $\VZr$. It is experimentally found that $\VO$ concentration increases significantly at the sample's surface layer~\cite{jo24} and grain boundary interfaces~\cite{peko13} after FS.
Because of its donor nature, such an avalanche of $\VO$ defects might be the origin of the high flash current when the flash event
occurs~\cite{jo24,lebrun17,raj12,vikrant20}.

It must be emphasized, however, that defect complexation in YSZ is fundamentally expected~\cite{nowick91}, driven by two coupled factors -- charge compensation and energy stabilization. Yet, the detailed atomic geometry of such complexes remains debated between theoretical reports~\cite{bogicevic01,lee11,guan20} (favoring $\VO\dchg{+2}$ as a second nearest neighbor to $\YZr\dchg{-1}$, as illustrated in Fig.~\ref{fig:2}h,i) and experimental interpretations. Nevertheless, experimental evidence supports the existence of specific geometry configurations which are typically modeled as $\YZr\VO\YZr$ ``trimer''~\cite{wang00} and $\YZr\VO$ pair~\cite{paiverneker89}. We thus calculate the formation energies of the $\trimer\dchg{q}$ and $\pair\dchg{q}$ complexes,
Fig.~\ref{fig:1}. The $\YZr\VO\YZr$ trimer (Fig.~\ref{fig:2}h) remains in neutral state by
balancing the charge of $\VO\dchg{+2}$ with two $\YZr\dchg{-1}$ for $\Ef > \varepsilon(+1/0) = 0.43$~eV, which significantly affects the defect equilibrium in YSZ as will be shown later. Instead of being exclusively neutral as usually assumed, it can
be in $+1$ charge state as a deep donor for $\Ef < 0.43$~eV. The $\YZr\VO$ pair (Fig.~\ref{fig:2}i) is in $+1$ charge state for $\Ef < \varepsilon(+1/-1) = 3.56$~eV, effectively representing $\YZr\dchg{-1} + \VO\dchg{+2}$ association.

The kinetics of FS is determined by the migration of heavier Zr atoms. Among the Zr defects, the higher formation energies of Zr interstitials ($\Zri$) and Zr anti-sites ($\ZrO$) than Zr vacancy ($\VZr$) make them less important in FS (see SI, Fig.~S3). We
hereafter focus on $\VZr\dchg{q}$ that can accept up to four electrons, $q = 0, -1, -2, -3, -4$, in four-valent ZrO$_2$. Such vacancies form by O incorporation
from the gas, O$_2$(g) $\to$ 2O$_\text{O}^{\times} + \VZr\dchg{0}$ (O$_\text{O}^{\times}$ denoting lattice oxygen; cf.\ SI, Sec.~S3), which conserves the
number of Zr atoms~\cite{smyth00} and the prevailing charge state is then set by \Ef\ 
(Fig.~\ref{fig:1}). The formation of $\VZr$ breaks eight Zr-O bonds (Fig.~\ref{fig:2}e),
resulting in eight O dangling bonds. The neutral $\VZr\dchg{0}$ leaves four holes that are
derived mostly from O~$2p$ states and shown as the two in-gap peaks (Fig.~\ref{fig:2}b);
the holes prefer to spatially localize on four individual O atoms next to $\VZr$ in a tetrahedral coordination to minimize Coulomb repulsion and form small hole polarons. Such small polarons have also been reported for $\VZr\dchg{0}$ in its polymorph isovalent monoclinic HfO$_2$/ZrO$_2$~\cite{mckenna12,osorio07}. 
The small hole polarons, induced by the presence of 
unpaired spin-polarized electrons as illustrated in
Fig.~\ref{fig:2}e, are stabilized by lattice distortions near the defects and give rise to
magnetic moment $\sim 0.76\mu_\text{B}$ on each O atom, which distort towards the $\VZr$ void and
elongates the nearby Zr-O bonds to 2.35~\AA, compared to 2.22~\AA\ in pristine ZrO$_2$.

\begin{figure*}[t]
  \includegraphics[width=0.8\textwidth]{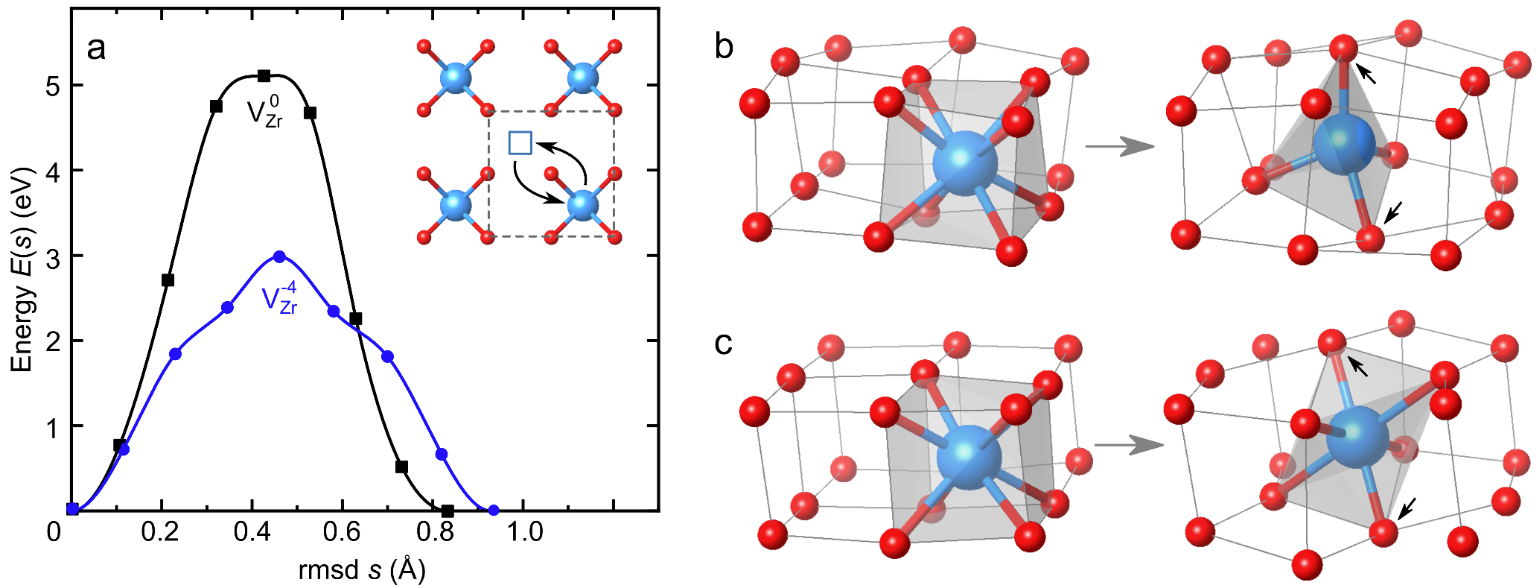}
  \caption{\label{fig:3}%
    (a) Potential energy profile $E(s)$ of Zr from its lattice site to a nearby Zr vacancy
    $\VZr$ (inset) in a neutral (0) and $(-4)$-charge states (as labeled) versus the
    cumulative root mean square displacement (rmsd)
    $s_i = s_{i-1} + \bigl(\sum_k^N \|\mathbf{R}_k^i - \mathbf{R}_k^{i-1}\|^2/N\bigr)^{1/2}$.
    The inset illustrates the Zr-ion $\rightleftharpoons$ $\VZr$ swap (view along [001]).
    The large dashed square indicates the section shown in (b) and (c); the $\VZr$ vacancy
    is shown as $\square$. Local geometry along the migration pathway for (b)
    $\VZr\dchg{0}$ and (c) $\VZr\dchg{-4}$ (Zr: blue; O: red) in the initial (left) and
    transition (right) states. Local coordination of Zr is highlighted (shaded polyhedra).
    Small arrows indicate a pair of O atoms at the TS undergoing large displacements.}
\end{figure*}

Singly charged $\VZr\dchg{-1}$ is always metastable with higher formation energy than other charge states for any \Ef. $\VZr$ can also adopt doubly ($\VZr\dchg{-2}$) or triply ($\VZr\dchg{-3}$) negative charge states if $\Ef$ is within 1.61--1.72~eV or 1.72--1.78~eV, respectively. The hole polaron states in $\VZr\dchg{0/-1/-2/-3}$ result in
the paramagnetic nature of vacancies, rendering them observable through electron spin/paramagnetic resonance (ESR/EPR) spectroscopy~\cite{emanuelsson93}. In quadruple
negative $\VZr\dchg{-4}$, localized hole states become fully occupied and lie as resonances in the valence band. A small polaron usually results in a low electrical (hopping) conductivity~\cite{peng17}, hence an abrupt rise of the conductivity at the onset of FS can point to a $(0/-4)$ charge state transition of $\VZr\dchg{q}$, ensued by the disapperance of the small polaron. The $\VZr$ charge transition level $\varepsilon(0/-4) \simeq 1.78$~eV indicates that it is a deep acceptor and $\VZr\dchg{-4}$ will be favorable, with a higher concentration, if $\Ef$ shifts closer to CBM during flash.

\textit{Charge-state dependent cation migration barrier}---Given the importance of volume diffusion of defects for mass transport in FS, we calculate the charge state-dependent migration barriers of $\VZr\dchg{q}$ for $q = 0, -4$, as shown in Fig.~\ref{fig:3}a. The migration process can be considered as a nearest-neighbor ion~$\rightleftharpoons$~vacancy, Zr$^{4+} \rightleftharpoons \VZr$, position swapping along the $\langle 110 \rangle$ directions. Interestingly, the migration barrier for a neutral vacancy is $\Delta E(\VZr\dchg{0}) = 5.10$~eV, more than 2~eV higher than that for a $-4$ charge state, $\Delta E(\VZr\dchg{-4}) = 2.98$~eV.

The origin of such a large ($\simeq 40\%$) barrier reduction is examined from two aspects. The larger Zr-O bond distortion towards the $\VZr\dchg{0}$ void, induced by the small polarons, results in reduced free space in $\VZr\dchg{0}$ for the migrating Zr to
accommodate its coordination with nearby O. As indicated in Fig.~\ref{fig:3}b,c, at the
transition state the migrating Zr is tetrahedrally coordinated by O for $\VZr\dchg{0}$,
whereas for $\VZr\dchg{-4}$ it favors a more stable octahedral coordination that closely
resembles the eight-coordinated cube in the original lattice. Besides, small polaron
hopping in $\VZr\dchg{0}$ between the initial and final positions synchronizes with the
cation migration but goes in the opposite direction. Roughly 75\% of the four hole polarons
of $\VZr\dchg{0}$ are relocated when migrating to the transition state (see SI, Fig.~S4 and
S5), such hopping may also contribute to the higher migration barrier of $\VZr\dchg{0}$.

Based on the computed vacancy migration barrier $\Delta E$, we estimate the onset temperature $T^*$ at which $\VZr$ defects become mobile by taking the jump rate $\Gamma = \Gamma_0 \exp(-\Delta E/\kB T^*) \sim 1$~s$^{-1}$, where the prefactor is approximated as a typical phonon frequency $\Gamma_0 \sim 10^{13}$~s$^{-1}$~\cite{souvatzis08}, and $\kB$ is the Boltzmann constant. The estimated $T^* \sim 1200$~K for $\VZr\dchg{-4}$ is in good agreement with experimental observation of FS temperature~\cite{biesuz19}. 
It is interesting to note that the experimentally observed enhanced M$^{4+}$ cation mobility in Zr- and Ce-containing ceramics hypothetically arises from a lower diffusion barrier for the reduced cation (effectively M$^{3+}$, M = Zr, Ce)~\cite{dong22} -- a transient complex M$^{4+}$-$e^-$, ``polaronium'', of a migrating M$^{4+}$ cation dressed with a small
electron polaron ($e^-$). A favorable formation of a $\VZr\dchg{-4}$Zr$^{4+}$-$e^-$ vacancy-polaronium complex requires its binding energy $E_\text{b} = \Eform(\VZr\dchg{-4}) + \Eform(\mathrm{Zr}^{4+}\text{-}e^-)
- \Eform(\VZr\dchg{-4}\mathrm{Zr}^{4+}\text{-}e^-) > 0$. We find, however, that $E_\text{b} \simeq -1.05$~eV due to the strong Coulomb repulsion between negatively charged $\VZr\dchg{-4}$ and the self-trapped $e^-$ at a nearby Zr$^{4+}$ and do not further consider such a polaronium mechanism.

The electron-capture behavior of $\VZr$ leading to the $\VZr\dchg{-4}$ state resembles a two-step charge ``ratchet'', by virtue of the $(-1/-2)$ and $(-3/-4)$ transitions. The electron capture coefficients of $\VZr$ (see SI, Table~S1 and Sec.~S4) are relatively low due to the large bandgap and its deep acceptor transition level, which contributes to the demanding process conditions (high $T$ and electric field) for FS of ZrO$_2$.

\begin{figure}[t]
  \includegraphics[width=\columnwidth]{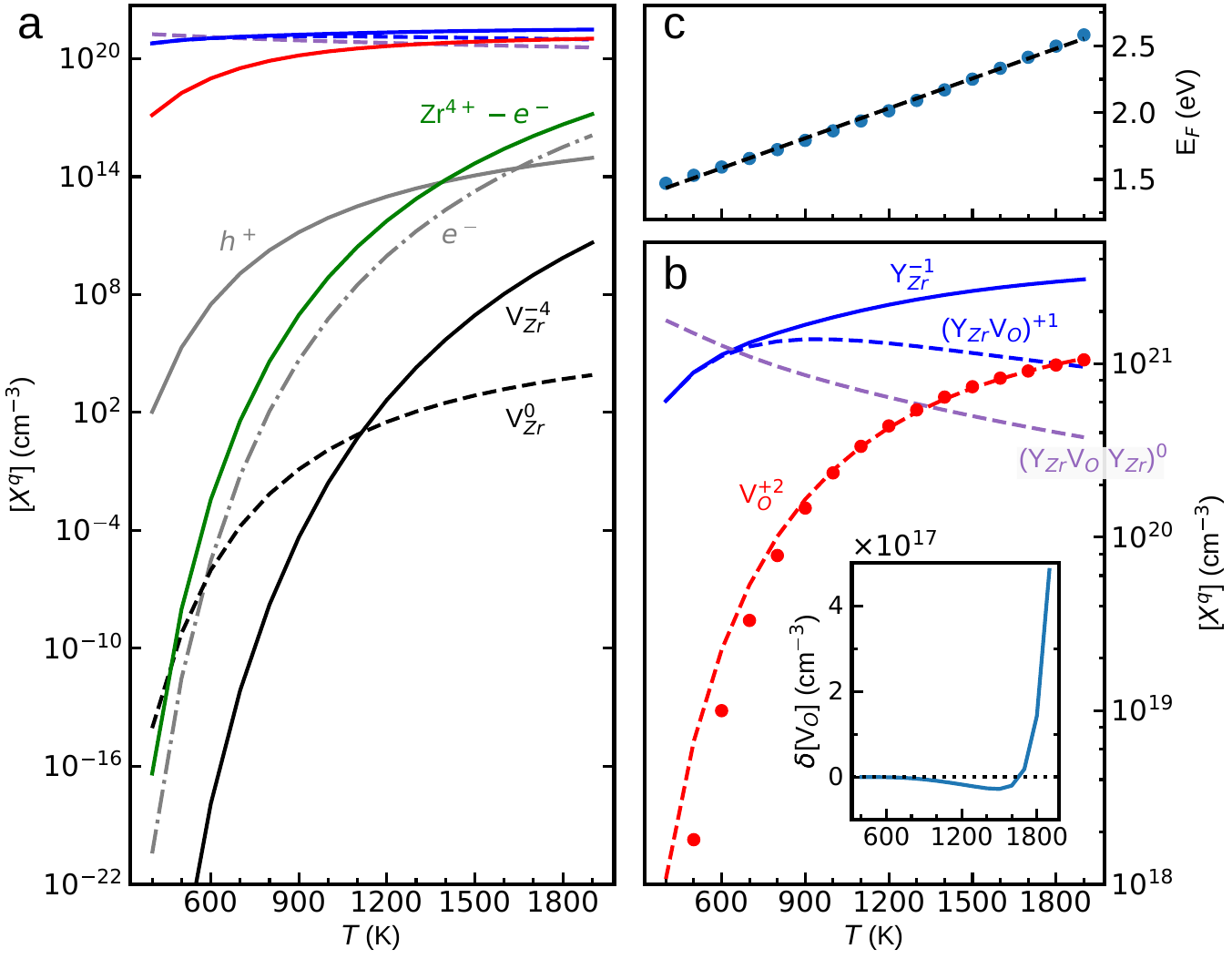}
  \caption{\label{fig:4}%
    Computed temperature dependences of (a,b) defect concentrations $[X\dchg{q}]$ and
    (c) Fermi level \Ef. Zoom-in of the concentration range of the dominant defects in
    (a) $\{\VO\dchg{+2}, \YZr\dchg{-1}, \pair\dchg{+1}, \trimer\dchg{0}\}$ is shown in (b).
    For $[\VO\dchg{+2}]$, the actually computed values are shown as points ($\bullet$) and
    the red dashed line is the fit to Eq.~(S16) (see SI, Sec.~S3) resulting in an effective
    $\varepsilon \simeq 0.63$~eV. The inset in (b) shows the excess O-vacancy concentration
    $\delta[\VO] = [\VO]_\text{total} - \tfrac{1}{2}[\text{Y}]_\text{total}$ as a
    function of temperature. In (c) the points are the computed values and the dashed line
    is the linear regression using Eq.~(S19) (see SI, Sec.~S3).}
\end{figure}

\textit{Defect equilibria and the Fermi level of YSZ in flash sintering}---To understand
how defect concentrations and their dominant charge state evolve during FS, $\Ef$ and
$[X\dchg{q}]$ are determined self-consistently as shown in Fig.~\ref{fig:4}, assuming an
ensemble of various point defects in the dilute limit with negligible defect interactions
(see SI, Secs.~S1 and S2).

As can be seen in Fig.~\ref{fig:4}a, the defect chemistry in YSZ is dominated by the individual ($\YZr$, $\VO$) and complexated ($\YZr\VO$, $\YZr\VO\YZr$) defects across all temperatures. Although the system is considered thermodynamically open, in contact with the O$_2$ gas environment at ambient pressure, concentration of the dominant free/mobile species $[\VO\dchg{+2}]$ can be accurately described (Fig.~\ref{fig:4}b) by considering a lattice-gas model (for details, see SI, Section~S3) of an effectively closed system of
$\VO$ that can bind/adsorb on $\YZr$ sites with a binding energy $\varepsilon$. Such an effectively closed model simplifies the defect equilibria to a quadratic equation for the free $\VO$. However, with rising temperature and increasing $[h^+]$, a small fraction of excess vacancies $\delta[\VO]$ are generated (inset in Fig.~\ref{fig:4}b), that drives the nearly linear upshift of the Fermi energy with temperature. This leading $\Ef \sim T$ behavior, Fig.~\ref{fig:4}c, is governed by the weak oxidation, $\delta[\VO] < 0$ up to $T \lesssim 1670$~K, where the system inhales O from the gas environment to maintain charge balance, $2[\VO\dchg{+2}] + [h^+] = [\text{Y}^{-1}] \equiv \text{const}$, yielding
$\Ef(T) \simeq \Delta H_{\mathrm{eff}} - T\Delta S_{\mathrm{eff}}$ with an effective
$\Delta H_{\text{eff}} = 1.14$~eV and entropy loss
$\Delta S_{\text{eff}} = -0.75$~meV~K$^{-1}$, whose magnitude is compatible with an estimate based on standard-state O$_2$ gas-phase thermochemistry (for details see SI, Sec.~S3).

During FS, the temperature evolution $T(t)$ of an electrically loaded YSZ sample as a blackbody with uniform $T$ is given by the heat transport equation~\cite{vikrant20,dong15} which, for a cubic sample of side $L$, can be written as:
\begin{equation}
  \rho C \frac{dT}{dt}
  = \frac{e^2 q^2 [\VO\dchg{q}]\, D_{\VO}}{\kB T}\,\mathcal{E}^2 - \frac{6\epsilon\sigma_e}{L} \left(T^4 - T_\text{f}^4\right),
  \label{eq:4}
\end{equation}
where $\rho$ and $C$ are the sample density and specific heat capacity, $\mathcal{E}$ is
the applied electric field, $e$ elementary charge, $q$ and $D_{\VO}$ are the charge and
diffusion coefficient of $\VO$, $\epsilon$ the emissivity, $\sigma_e$ the Stefan--Boltzmann constant, and $T_\text{f}$ the furnace temperature. The first term on the right-hand side represents the Joule heating mainly due to the mobile $\VO$ (dominantly in $q = +2$ charge state in the relevant range of \Ef, Fig.~\ref{fig:1}), and the second --
the heat radiation losses (see SI, Sec.~S5 for more details).

The fact that YSZ samples exhibit low conductivity in stage I implies that oxygen vacancies
may exist mostly in the form of immobile $\YZr\VO\YZr$ complexes. Indeed, we find that for
$T \lesssim 600$~K, $\trimer\dchg{0}$ is the dominant defect as seen in Fig.~\ref{fig:4}b. When a 8YSZ sample is heated in a furnace during stage I, the weak oxidation (inset, Fig.~\ref{fig:4}b) inevitably backfills some bound $\VO$ thus annihilating a small amount of $\YZr\VO\YZr$ complexes, which ``release'' $\YZr\dchg{-1}$. Oxygen vacancies still remain
mostly trapped in $\pair\dchg{+1}$ pairs with $[\pair\dchg{+1}]$ rising until $T \sim 800$--1000~K. It is interesting to note, however, that in this incubation stage, $T \lesssim 1000$~K, the electronic carrier population is dominated by holes
$[h^+] \gg [e^-]$, Fig.~\ref{fig:4}a, in agreement with the experimentally observed $p$-type conductivity~\cite{vendrell19,jo20,dong21}, and arising from the slight sample oxidation to maintain the equilibrium in the ambient atmosphere, as mentioned above.

\begin{figure}[!tb]
  \includegraphics[width=\columnwidth]{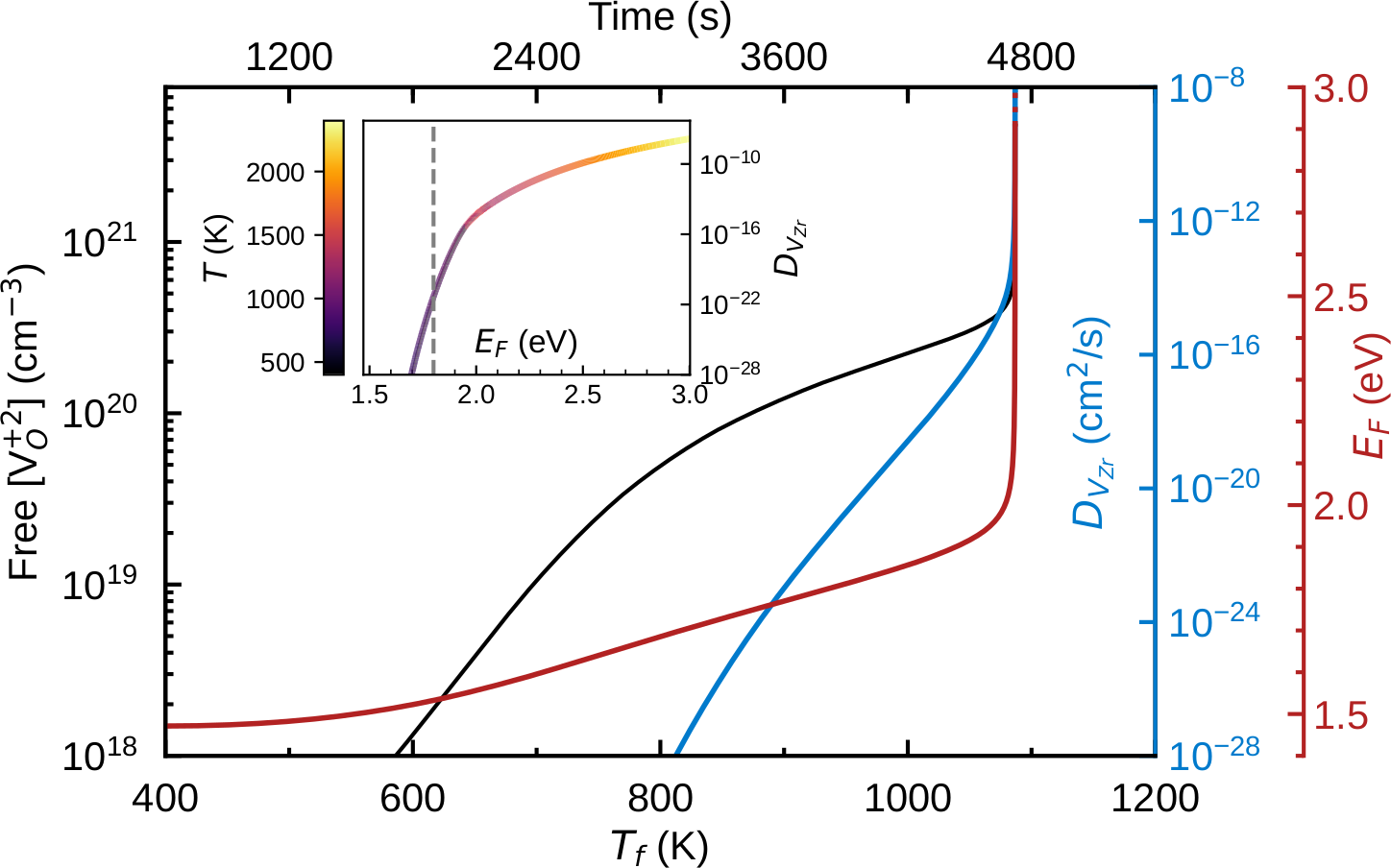}
  \caption{\label{fig:5}%
    Diffusion coefficient of $\VZr$, $D_{\VZr}$, Fermi level \Ef\ (right axes), and the
    concentration of free oxygen vacancies $\VO\dchg{+2}$ (left axis) as a function of
    furnace temperature $T_\text{f}(t) = T_0 + \kappa t$, with $T_0 = 300$~K and
    $\kappa = 10$~K/min. The inset shows $D_{\VZr}$ as a function of \Ef\ with color coding
    indicating the sample temperature $T$. The vertical dashed line marks the
    $\varepsilon(0/-4)$ charge transition level of $\VZr$ (cf.\ Fig.~\ref{fig:1}).}
\end{figure}

Subsequently, at the onset of stage II, the Joule heating (thermal runaway) triggers dissociation of both defect complexes, the dominant charged pair and the neutral trimer, to individual $\VO\dchg{+2}$ and $\YZr\dchg{-1}$, the concentration of mobile and conductive doubly ionized $\VO\dchg{+2}$ escalates abruptly as shown in Fig.~\ref{fig:5}, which raises
the sample temperature-$T$/current instantaneously and possibly relates to the defect avalanche~\cite{cologna10,jo24}. Beyond $T \sim 1700$~K, the system is reduced as evident by the relatively small positive excess $\delta[\VO] > 0$ in the inset of Fig.~\ref{fig:4}b. Therefore, the weak-oxidation rationale for the nearly linear $\Ef(T)$
no longer applies above this crossover. Nevertheless, $\Ef$ continues to rise because the excess oxygen vacancies $\delta[\VO\dchg{+2}]$ release electrons, thus the electroneutrality is satisfied at a higher $\Ef \sim 2$--2.5~eV (Fig.~\ref{fig:4}c). Such oxygen deficiency has recently been experimentally observed in the surface layer of a single-crystal cubic ZrO$_2$ sample~\cite{jo24}.

The continuously rising Fermi level, as summarized in Fig.~\ref{fig:5}, throughout the FS governs the concentration of $\VZr$ that varies with the charge state. As shown in Fig.~\ref{fig:4}a, $[\VZr\dchg{0}] \gg [\VZr\dchg{-4}]$ when the Fermi level is lower in
stage I, whereas $[\VZr\dchg{-4}]$ surges remarkably and $\VZr\dchg{-4}$ becomes the dominant charge state of Zr vacancies when the Fermi level is lifted up in stage II. The effective (mean) tracer diffusion coefficient of $\VZr$ is evaluated in Fig.~\ref{fig:5} as $D_{\VZr} \equiv \langle D^q_{\VZr} \rangle_q = 2a^2 \sum_q p_q \Gamma_q$, where $a$ is the Zr$^{4+}$/$\VZr$ migration distance,
$p_q = [\VZr\dchg{q}]/([\VZr\dchg{0}] + [\VZr\dchg{-4}])$ is the probability of being in
charge-$q$ state ($q = -4$ or 0) and $\Gamma_q$ is the corresponding jump rate. At 1900~K, the cation diffusion length over $t \sim 1$~s is $\sim (D_{\VZr} t)^{1/2} = 0.16$~$\mu$m, which is of similar order of magnitude as the grain size in YSZ powders~\cite{yadav17}.

In a nutshell, a plausible microscopic mechanism of enhanced cation mobility in FS can be outlined: in the course of FS, O$_2$ inhalation and exhalation of the heated YSZ sample annihilate and create electron donors $\VO$, whereby Fermi level is lifted up continually from a lower position that favors $\VZr\dchg{0}$ to a higher one that supports
$\VZr\dchg{-4}$, thus $\VZr$ migrates with a much lower energy barrier, enhancing sintering kinetics.

\textit{Discussion}---Given that defects and the small polaron localized nearby are common
in metal oxides~\cite{zunger21,lany09}, we have also examined the charge-state-dependent migration barriers $\Delta E(\VM\dchg{q}, q = 0, -4)$ of cation
vacancies $\VM$ in other oxide ceramics of practical interest, such as monoclinic ZrO$_2$ and rutile TiO$_2$, that can be similarly flash sintered~\cite{biesuz19}. The barriers are
lowered by 1.29~eV and 1.32~eV in $\VM\dchg{-4}$ (39\% and 32\%, as shown in SI, Fig.~S8),
respectively. Note that the mechanism of Fermi level shift in monoclinic ZrO$_2$ can be similar to the cubic phase we have investigated. It has been reported that vanadium- or nitrogen-doped rutile TiO$_2$ can be $p$-type~\cite{chang11,panepinto21} before FS and
excessive oxygen vacancies during FS induce $n$-type conductivity, such evolution in dominant defects would shift \Ef, trigger $\VTi$ charge transition and eventually lower $\VTi$ migration barrier, which provides a possible rationale for the observed lower flash temperature in vanadium- or nitrogen-doped rutile TiO$_2$ than in undoped one~\cite{zhang16}. Such a Fermi-level-mediated mechanism can thus be generalized to monoclinic ZrO$_2$ and rutile TiO$_2$. Quantities such as defect spin, carrier density and
specimen conductivity are generally accessible from experiments and can corroborate the present theoretical findings~\cite{emanuelsson93}.

Another important mechanism stimulating sintering is the grain boundary
migration~\cite{klein23,zou19,zou15,xu21}, which involves space-charge layers comprising a charged grain boundary core and a layer of oppositely charged point defects. Consequently, the grain boundary migration would be enhanced, if cation vacancies prevail with low migration barriers along the network of grain boundaries~\cite{klein23,xu21}, or if the migration dynamics of the grain-boundary core can be tailored significantly by chemical potentials of species and electron doping in its constituent
dislocations~\cite{zou19,zou15}.

In summary, our extensive first-principles calculations of point defects in YSZ reveal an
atomistic mechanism for the enhanced Zr$^{4+}$ cation migration associated with defect
charge-state transition. At the incubation stage of FS, effectively all $\VO$ are buffered
in $\YZr\VO\YZr$ and $\YZr\VO$ complexes while the weak sample oxidation leads to extrinsic
electronic $p$-type leakage; \Ef\ is closer to the valence band and favors $\VZr\dchg{0}$.
As the flash event initiates, a $[\VO]$-surge originates from dissociation of the charged
$\pair\dchg{+1}$ mainly, as well as the neutral $\YZr\VO\YZr$ complexes resulting in thermal
runaway. The sharp rise in $T$ triggers thermal reduction of the sample, with the excess
non-stoichiometric $\VO\dchg{+2}$ contributing thermally excited electrons to form self-trapped electron polaron on lattice Zr and to the conduction band edge. This lifts up \Ef\ higher where a more mobile $\VZr\dchg{-4}$ becomes the dominant charge state of Zr vacancies. The $\simeq 2$~eV ($\simeq 40\%$ lowering) cation migration barrier reduction
triggered by the Fermi-level-mediated charge-state transition presents a plausible and generalizable microscopic cause of the accelerated mass transport in FS of metal oxide ceramics.

\textit{Supporting Information} -- The Supporting Information is available free of charge
at \dots\ 

Computational method; Determination of chemical potentials; \Ef\ and defect
concentrations; Temperature dependence of \Ef; $\VZr\dchg{q}$ charge-state transitions by
non-radiative carrier capture; Solving the heat transport equation.
\medskip

The authors declare no competing financial interest.
\medskip

\begin{acknowledgments}
\textit{Acknowledgments}---This work was supported by the Office of Naval Research (ONR), Grant N00014-22-1-2788. Computer resources were provided through DOE's NERSC award BES-ERCAP0037090, and allocation DMR100029 from the NSF's ACCESS program.
\end{acknowledgments}

\bibliography{refs}

\end{document}